\documentstyle[12pt]{article}
\begin{document}
\topmargin -0.2cm \oddsidemargin -0.2cm \evensidemargin -1cm
\textheight 22cm \textwidth 12cm

\title{Two new type surface polaritons excited into
nanoholes in metal films.}
\author{Minasyan V.N. and Samoilov V.N.\\
Scientific Center of Applied Research, JINR,\\
 Dubna, 141980, Russia}

\date{\today}

\maketitle

\begin{abstract}
First, we argue that the smooth metal-air interface should be regarded as a distinct dielectric medium, the skin of the metal. The existence of this metal skin leads to theoretical explanation of experimental data on the excitation of electromagnetic surface shape resonances in lamellar metallic gratings by light in the visible to near-infrared range. Surface polaritons have been observed in reflection modes on metallized gratings where the electric field is highly localized inside the grooves (around 300-1000 times larger than intensity of incoming optical light). Here we present quantized Maxwell's equations for electromagnetic field in an isotropic homogeneous medium, allowing us to solve the absorption anomaly property of these metal films. The results imply the existence of light boson quasi-particles with spin one and mass$m= 2.5\cdot 10^{-5} m_e$. We also show the presence of two new type surface polaritons into nanoholes in metal films.
\end{abstract}

\vspace{100mm}

\vspace{5mm}

{\bf 1. Introduction.}

\vspace{5mm}

There have been many studies of optical light transmission through individual nanometer-sized holes in opaque metal films in recent years [1-3]. These experiments showed highly unusual transmission properties of metal films perforated with a periodic array of subwavelength holes, because the electric field is highly localized inside the grooves (around 300-1000 times larger than intensity of incoming optical light). Here we analyze the absorption anomalies for light in the visible to near-infrared range observed into nanoholes in metal films. These absorption anomalies for optical light as seen as enhanced transmission of optical light in metal films, and attributed to surface plasmons (collective electron density waves propagating along the surface of the metal films) excited by light incident on the hole array [4]. The enhanced transmission of optical light is then associated with surface plasmon (SP) polaritons. Clearly, the definition of surface metal-air region is very important factor, since this is where the surface plasmons are excited. In contrast to this surface plasmon theory, in which the central role is played by collective electron density waves propagating along the surface of metal films in a free electron gas model, the authors of paper [5] propose that the surface metal-air medium should be regarded as a metal skin and that the ideas of the Richardson-Dushman effect of thermionic emission are crucial [6]. Some of the negatively charged electrons are thermally excited from the metal, and these evaporated electrons are attracted by positively charged lattice of metal to form a layer at the metal-air interface. However, it is easy to show that the thermal Richardson-Dushman effect is insufficient at room temperature $T\simeq 300K$ because the exponent $ \exp^{-\frac{\phi}{k T}} $  with a value of the work function $\phi \simeq 1eV-10eV$  leads to negligible numbers of such electrons.
In this letter, we shall regard the metal skin as a distinct dielectric medium consisting of neutral molecules at the metal surface. Each molecule is considered as a system consisting of an electron coupled to an ion, creating of dipole. The electron and ion are linked by a spring which in turn defines the frequency $\omega_0$ of electron oscillation in the dipole. Obviously, such dipoles are discussed within elementary dispersion theory [7]. Further, we shall examine the quantization scheme for local electromagnetic field in the vacuum, as first presented by Planck for in his black body radiation studies. In this context, the classic Maxwell equations lead to appearance of the so-called ultraviolet catastrophe; to remove this problem, Planck proposed modelled the electromagnetic field as an ideal Bose gas of massless photons with spin one. However, Dirac [8] showed the Planck photon-gas could be obtained through a quantization scheme for the local electromagnetic field, presenting a theoretical description of the quantization of the local electromagnetic field in vacuum by use of a model Bose-gas of local plane electromagnetic waves, propagated by speed c in vacuum. An investigation of quantization scheme for the local electromagnetic field [9] predicted the existence of light quasi-particles with spin one and finite effective mass $m= 2.5\cdot 10^{-5} m_e$ (where $m_e$ is the mass of electron) by introducing quantized Maxwell equations. In this letter, we present properties of photons which are excited in clearly dielectric medium, and we show existence of two new type surface polaritons into nanoholes in metal films.

\vspace{5mm}

{\bf 11. Quantized Maxwell equations.}

\vspace{5mm}

We now investigate Maxwells equations for dielectric medium [7] by quantum theory field [8]:

\begin{equation}
curl {\vec {H}} -\frac{1}{c}\frac{d {\vec{D}}}{d t}=0
\end{equation}

\begin{equation}
curl {\vec {E}} +\frac{1}{c}\frac{d {\vec{B}}}{d t}=0
\end{equation}

\begin{equation}
div {\vec {D}} =0
\end{equation}

\begin{equation}
div {\vec {B}} =0
\end{equation}

where $\vec {B}=\vec {B}(\vec {r},t)$ and $\vec {D}=\vec {D}(\vec {r},t)$ are, respectively, the local magnetic and electric induction depending on space coordinate $\vec {r}$ and time $t$; $\vec {H}=\vec {H}(\vec {r},t)$ and $\vec {E}=\vec {E}(\vec {r},t)$ are, respectively, the magnetic and electric field vectors, and $c$ is the velocity of light in vacuum. The further equations are
\begin{equation}
\vec {D} =\varepsilon \vec {E}
\end{equation}

\begin{equation}
\vec {B} =\mu \vec {H}
\end{equation}

where $\varepsilon >1$ and $\mu=1$ are, respectively, the dielectric and the magnetic susceptibilities of the dielectric medium.

The Hamiltonian of the radiation field $\hat{H}_R$ is:

\begin{equation}
\hat{H}_R =\frac{1}{8\pi}\int \biggl(\varepsilon E^2+\mu H^2\biggl) dV
\end{equation}

We now wish to solve a problem connected with a quantized electromagnetic field, and begin from the quantized equations of Maxwell. We search for a solution of (1)-(6), in an analogous manner to that  presented in [9]:

\begin{equation}
\vec {E}=- \frac{\alpha}{c}\cdot
\frac{d {\vec{H}_0}}{d t}+\beta\cdot \vec {E}_0
\end{equation}
and
\begin{equation}
\vec {H}= \alpha \cdot  curl{
\vec {H}_0} + \beta \vec{H}_0
\end{equation}
where $\alpha=\frac{\hbar \sqrt{2\pi}}{\sqrt{m}}$ and $\beta= c\sqrt{2m\pi} $   are the constants obtained in [9]. Thus $\vec {E}_0=\vec {E}_0(\vec {r},t)$ and $\vec {H}_0=\vec {H}_0 (\vec {r},t)$ are, respectively, vectors of electric and magnetic field for one Bose-light-particle of electromagnetic field with spin one and finite effective mass $m$. The vectors of local electric $\vec {E}_0$ and magnetic $\vec {H}_0$ fields, presented by equations (8) and (9), satisfy to equations of Maxwell in dielectric medium:

\begin{equation}
curl {\vec {H}_0} -\frac{\varepsilon }{c}\frac{d {\vec{E}_0}}{d t}=0
\end{equation}

\begin{equation}
curl {\vec {E}_0} +\frac{1}{c}\frac{d {\vec{H}_0}}{d t}=0
\end{equation}

\begin{equation}
div {\vec {E}_0} =0
\end{equation}

\begin{equation}
div {\vec {H}_0} =0
\end{equation}

By using of (10),  we can rewrite (9) as
\begin{equation}
\vec {H}= \frac{\alpha\varepsilon }{c}\frac{d {\vec{E}_0}}{d t} + \beta \vec{H}_0
\end{equation}

The equations (10)-(13) lead to a following wave-equations:
\begin{equation}
\nabla^2 {\vec {E}_0}-\frac{\varepsilon }{c^2}\frac{d^2 \vec{E}_0}{d t^2}=0
\end{equation}
and
\begin{equation}
\nabla^2 {\vec {H}_0}-\frac{\varepsilon }{c^2}\frac{d^2 \vec{H}_0}{d t^2}=0
\end{equation}

which in turn have the following solutions:
\begin{equation}
\vec {E}_0= \frac{1}{\sqrt{V}}\sum_{\vec{k}}\biggl(
\vec {E} _{\vec{k}} e^{i(
\vec{k}\vec{r} + \frac{kc t}{\sqrt{\varepsilon }})} +\vec {E}^{+}_{\vec{k}}
e^{-i(\vec{k}\vec{r} + \frac{kc t}{\sqrt{\varepsilon }})}\biggl)
\end{equation}

\begin{equation}
\vec {H}_0= \frac{1}{\sqrt{V}}\sum_{\vec{k}}\biggl(
\vec {H} _{\vec{k}} e^{i(
\vec{k}\vec{r} + \frac{kc t}{\sqrt{\varepsilon }})} +\vec {H}^{+}_{\vec{k}}
e^{-i(\vec{k}\vec{r} + \frac{kc t}{\sqrt{\varepsilon }})}\biggl)
\end{equation}
where  $\vec { E } ^{+}_{\vec{k}}$, $\vec { H } ^{+}_{\vec{k}}$  and  $\vec {E} _{\vec{k}}$, $\vec {H} _{\vec{k}}$ are, respectively, the second quantzation vector wave functions, essentially the vector Bose "creation" and "annihilation" operators  for the Bose quasi-particles of electric and magnetic waves with spin one in dielectric medium. With these new terms $\vec {E}_0$ and $\vec {H}_0$, the radiation Hamiltonian $\hat{H}_R$ in (7) takes the form:

\begin{equation}
\begin{array}{ll}
\hat{H}_R =\frac{1}{8\pi }\int \biggl(\varepsilon E^2+H^2\biggl) dV =\\[+12pt]
\displaystyle
=\;\frac{1}{8\pi}\int \biggl [\varepsilon \biggl(- \frac{\alpha}{c}\frac{d {\vec{H}_0}}{d t}+\beta \vec {E}_0\biggl)^2+\\[+12pt]
\displaystyle
+\;\biggl(\frac{\alpha\varepsilon }{c}\frac{d {\vec{E}_0}}{d t}+ \beta \vec {H}_0\biggl)^2\biggl] dV
\end{array}
\end{equation}

where, by substituting into (17) and (18), leads to the reduced form of $\hat{H}_R$, at condition of transverse electromagnetic wave
$$
\vec{E}_{\vec{k}}\cdot\vec{H}_{\vec{k}}=0
$$
$$
\vec{E}^{+}_{\vec{k}}\cdot\vec{H}_{\vec{k}}=0
$$
and
$$
\frac{1}{V} \int e^{i\vec{k}\cdot\vec{r}}dV=\delta_{\vec{k}}
$$

\begin{equation}
\hat{H}_R=\hat{H}_e+\hat{H}_h
\end{equation}

where the operators $\hat{H}_e $ and $\hat{H}_h$ are:

\begin{equation}
\begin{array}{ll}
\hat{H}_e =\sum_{\vec{k}}\biggl (\frac{\hbar^2
k^2 \varepsilon ^2}{2m }+
\frac{mc^2\varepsilon }{2}\biggl )
\vec{E}^{+}_{\vec{k}}\vec{E}_{\vec{k}}-\\[+12pt]
\displaystyle
-\;\frac{1}{2}\sum_{\vec{k}}\biggl(\frac{\hbar^2k^2\varepsilon ^2 }{2m }-
\frac{mc^2\varepsilon }{2}\biggl)
\biggl (\vec{E}^{+}_{\vec{k}}
\vec{E}^{+}_{-\vec{k}}+
\vec{E}_{-\vec{k}}\vec{E}_{\vec{k}}\biggl)
\end{array}
\end{equation}

and

\begin{equation}
\begin{array}{ll}
\hat{H}_h =\sum_{\vec{k}}\biggl (\frac{\hbar^2
k^2\varepsilon}{2m }+
\frac{mc^2}{2}\biggl )
\vec{H}^{+}_{\vec{k}}\vec{H}_{\vec{k}}-\\[+12pt]
\displaystyle
-\;\frac{1}{2}\sum_{\vec{k}}\biggl(\frac{\hbar^2k^2 \varepsilon}
{2m}-\frac{mc^2}{2}\biggl)
\biggl (\vec{H}^{+}_{\vec{k}}
\vec{H}^{+}_{-\vec{k}}+\\[+12pt]
\displaystyle
+\;\vec{H}_{-\vec{k}}\vec{H}_{\vec{k}}\biggl)
\end{array}
\end{equation}

In the letter [9], the boundary wave number $k_0=\frac{mc}{\hbar}$ for electromagnetic field in vacuum was appeared by suggestion that the light quasi-particles interact with each other by repulsive potential $U_{\vec{k}}$ in momentum space:
$$
U_{\vec{k}}=-\frac{\hbar^2k^2}
{2m}+\frac{mc^2 }{2}\geq 0
$$
As result, condition for wave numbers of light quasi-particles $k\leq k_0$ is appeared.

On other hand, due to changing energetic level into Hydrogen atom, the appearance of photon with energy $hkc$ is determined by a distance between energetic states for electron going from high level to low one. The ionization energy of the Hydrogen atom $E_I=\frac{m_e e^4}{2\hbar^2}$ is the maximal one for destruction atom. Therefore, one coincides with energy of free light quasi-particle $\frac{\hbar^2 k^2_0}{2m}$ which is maximal too because $k\leq k_0$. The later represents as radiated photon with energy $\hbar k_0 c $ in vacuum.  This reasoning claims the important condition as $\frac{m_e e^4}{2\hbar^2}=\hbar k_0 c$ which in turn determines a effective mass of the light quasi-particles $ m=\frac{m_e e^4}{2\hbar^2 c^2}=2.4 \cdot 10^{-35} kg $ in vacuum.

In analogy manner, we may find the boundary wave number $k_{\varepsilon}=\frac{mc}{\hbar\varepsilon}$ for light quasi-particles of electromagnetic field in isotropic homogenous medium by suggestion that light quasi-particles in medium interact with each other by repulsive potentials $U_{E,\vec{k}}$ in (21) and $U_{H,\vec{k}}$ in (22) which correspond, respectively, to electric and magnetic fields in momentum space:
$$
U_{E,\vec{k}}=-\frac{\hbar^2k^2 \varepsilon^2}
{2m}+\frac{mc^2\varepsilon }{2}\geq 0
$$
and

$$
U_{H,\vec{k}}=-\frac{\hbar^2k^2 \varepsilon}
{2m}+\frac{mc^2}{2}\geq 0
$$

Obviously, the both expressions in above determine wave numbers of light quasi-particles $k$ satisfying to condition $k\leq k_{\varepsilon}$.

We now apply a new linear transformation of the vector Bose-operators which is a similar to the Bogoliubov transformation [10] for scalar Bose operator, so as to evaluate the energy levels of the operator $\hat{H}_R$ within diagonal form:
\begin{equation}
\vec {E}_{\vec{k}}=\frac{\vec {e}_{\vec{k}} +
M_{\vec{k}}\vec {e}^{+}_{-\vec{k}}} {\sqrt{1-M^2_{\vec{k}}}}
\end{equation}
and
\begin{equation}
\vec {H}_{\vec{k}}=\frac{\vec {h}_{\vec{k}} +
L_{\vec{k}}\vec {h}^{+}_{-\vec{k}}} {\sqrt{1-L^2_{\vec{k}}}}
\end{equation}

where $M_{\vec{k}}$ and $L_{\vec{k}}$ are the real symmetrical functions of  a wave vector $\vec{k}$.

The operator Hamiltonian $\hat{H}_R$ within using of a canonical transformation takes a following form:
\begin{equation}
\hat{H}_R=
\sum_{ k\leq k_{\varepsilon}}\chi_{\vec{k}}\vec {e}^{+}_{\vec{k}}
\vec {e}_{\vec{k}}+\sum_{ k\leq k_{\varepsilon}}\eta_{\vec{k}}\vec {h}^{+}_{\vec{k}}
\vec {h}_{\vec{k}}
\end{equation}

Hence, we infer that the Bose-operators $\vec {e}^{+}_{\vec{k}}$,
$\vec {e}_{\vec{k}}$ and $\vec {h}^{+}_{\vec{k}}$, $\vec
{h}_{\vec{k}}$ are, respectively, the vector of "creation" and the
vector of "annihilation" operators of two types of free photons with
energies

\begin{equation}
\begin{array}{ll}
\chi_{\vec{k}} =\varepsilon\sqrt{\biggl (\frac{\hbar^2k^2\varepsilon^2 }{2m}+\frac{mc^2\varepsilon }{2 }\biggl )^2-\biggl (\frac{\hbar^2k^2\varepsilon^2}{2m}-
\frac{mc^2\varepsilon}{2 }\biggl)^2}=\\[+12pt]
\displaystyle
=\;\hbar k v_e
\end{array}
\end{equation}
and

\begin{equation}
\begin{array}{ll}
\eta_{\vec{k}} =\sqrt{\biggl (\frac{\hbar^2k^2\varepsilon}{2m}+\frac{mc^2}{2 }\biggl )^2-\biggl (\frac{\hbar^2k^2\varepsilon}{2m}-
\frac{mc^2}{2 }\biggl)^2}=\\[+12pt]
\displaystyle
=\;\hbar k v_h
\end{array}
\end{equation}

where $v_e= c\varepsilon^{\frac{3}{2}}$ and $v_h= c\varepsilon^{\frac{1}{2}}$ are, respectively, velocities
of photons excited by the electric and the magnetic field. Thus, we predict the existence of two types photons excited in dielectric medium, with energies $\chi_{\vec{k}}=\hbar k c\varepsilon^{\frac{3}{2}}$ and $\eta_{\vec{k}}=\hbar k c\varepsilon^{\frac{1}{2}}$ that depend on the dielectric response of the
homogeneous medium $\varepsilon$. The velocities of the two new type photon
modes $v_e= c\varepsilon^{\frac{3}{2}}$ and $v_h= c\varepsilon^{\frac{1}{2}}$ are more than velocity
$c$ of photon in vacuum because $\varepsilon>1 $. Obviously, the phase
velocity of light is given by $v_p=\frac{ c}{\sqrt{\varepsilon}}$, contradicting
the results obtained for $v_e= c\varepsilon^{\frac{3}{2}} $ and $v_h= c\varepsilon^{\frac{1}{2}} $.
This is the source of the absorption anomalies in isotropic homogeneous media.

\vspace{5mm}

{\bf 111. Skin of metal on the boundary metal-air.}

\vspace{5mm}

A standard model of metal regards it as a gas of free electrons with negative charge
$-e$ in a box of volume $V$ , together with a background of lattice ions of opposite charge $e$ to preserve charge neutrality. For the boundary of this metal with the vacuum, we introduce the concept of a metal skin comprising free neutral molecules at the metal surface. The skin then has a thickness similar to the size of the molecule, a small number of Bohr diameter $a=\frac{2\hbar^2}{m e^2}=1\dot A$. We assume $N_0$ molecules per unit area is $N_0=\frac{3}{4\pi r^3}$ (where $r=\frac{a}{2}$ is the Bohr radius) which in turn determines the dielectric constant of metal's skin $\varepsilon$ under an electromagnetic field in the visible to near-infrared range with frequency $\omega\leq \omega_0$, by the well known formulae:

\begin{equation}
\varepsilon =1+ \frac{4\pi N_0 e^2}{m_e\biggl(\omega^2_0-\omega^2\biggl)}
\end{equation}
As we show in below, namely, the anomalies property of light is observed near resonance frequency $\omega_0$.

\vspace{5mm}

{\bf 1V. Two new type surface polaritons excited in metal films.}

\vspace{5mm}

We now show that presented theory explains the absorption anomalies such as enhanced transmission of optical light in metal films. We consider the subwavelength sized holes into metal films as cylindrical resonator with partly filled homogeneous medium [11]. The hole contains vacuum which has boundary with metals skin with width $a=10^{-4}\mu m$ but the grooves radius is $d=0.75 \mu m$ as experimental data [2]. The standing electromagnetic wave is excited by incoming light with frequency $\omega$ related to the frequency of cylindrical resonator $\omega$ by following system of dispersion equations:

\begin{equation}
\left.
\begin{array}{c}
\frac{ J_1(\frac{\omega d}{c})}{ J_0(\frac{\omega d}{c})}=
\frac{ J_1(\frac{\omega\sqrt{\varepsilon } d}{c})}{J_0(\frac{\omega\sqrt{\varepsilon } d}
{c})} \\
J_0(\frac{\omega\sqrt{\varepsilon }( d+a)}{c})=0
\end{array}
\right\}
\end{equation}

where $ J_0 (z)$ and $ J_1 (z)$, are, respectively, the Bessel functions of zero and one orders.

There is observed a shape resonance in lamellar metallic gratings when frequency $\omega$ of optical light in the visible to near-infrared range coincides with resonance frequency of dipole $\omega_0$ in metal's skin because the dielectric response is given by
$$
\lim_{\omega\rightarrow\omega_0}\varepsilon \rightarrow\infty
$$

Therefore, the energies of two types of surface polaritons tend to infinity. This result confirms that the electric field is highly localized inside the grooves because the energy of electric field inside the grooves is $300-1000$ times higher than energy incoming optical light in air $\chi_{\vec{k}}= \eta_{\vec{k}}=\hbar k c$ as $\varepsilon=1$ in air. Thus, we have shown the existence of two new type surface polaritons with energies $\chi_{\vec{k}}$ and $\eta_{\vec{k}}$ which are excited into nanoholes.

The resonance frequency of dipole $\omega_0$ in metal's skin is defined from (29), at condition $\varepsilon \rightarrow\infty$ in the metal skin, which is fulfilled at $\omega=\omega_0$. In turn, this leads to following equation:
\begin{equation}
J_1\biggl (\frac{\omega_0 d}{c}\biggl)=0
\end{equation}

because second equation in (29) is fulfilled automatically at condition $\varepsilon \rightarrow\infty$.

The equation (30) has a root $\omega_0 =\frac{3.8 c}{d}$ which in turn determines the resonance wavelength $\lambda_0=\frac{2\pi c}{\omega_0}=1.24\mu m$. This theoretical result is confirmed by experiment [2], where the zero-order transmission spectra were obtained with a Cary-5 spectrophotometer using of incoherent light sources with a wavelength range $0.2\leq \lambda \leq 3.3 \mu m$. Thus, the geometry of hole determines the transmission property of light into nanoholes.

In conclusion, we may say that the theory presented above confirms experimental results on metal films, and in turn solves the problem connected with the absorption anomalies in isotropic homogeneous media.

\vspace{15mm}

{\bf Acknowledgements}

\vspace{5mm}

We are particularly grateful to Professor Marshall Stoneham F R S
(London Centre for Nanotechnology, and Department of Physics and Astronomy
University College London, Gower Street, London WC1E 6BT, UK) for valuable scientific support and corrected English.

\end{document}